\documentclass[lettersize,journal]{IEEEtran}
\usepackage{amsmath,amsfonts}
\usepackage{algorithmic}
\usepackage{algorithm}
\usepackage{array}
\usepackage[caption=false,font=normalsize,labelfont=sf,textfont=sf]{subfig}
\usepackage{textcomp}
\usepackage{stfloats}
\usepackage{url}
\usepackage{verbatim}
\usepackage{graphicx}
\usepackage{cite}
\usepackage{multirow}
\usepackage{hyperref}
\usepackage{color}
\newcommand{\jiaxing}[1]{{\color{black}#1}}
\newcommand{\revise}[1]{{\color{black}#1}}
\begin{document}

\title{Multi-Atlas Brain Network Classification through Consistency Distillation and Complementary Information Fusion}

\author{Jiaxing Xu, Mengcheng Lan, Xia Dong, Kai He, Wei Zhang, Qingtian Bian, Yiping Ke
\thanks{Jiaxing Xu, Mengcheng Lan, Xia Dong, Qingtian Bian, and Yiping Ke are with the College of Computing and Data Science, Nanyang Technological University, Singapore, 639798 Singapore (e-mail:\url{JIAXING003@e.ntu.edu.sg}; \url{LANM0002@e.ntu.edu.sg}; \url{xia.dong@ntu.edu.sg}; \url{BIAN0027@e.ntu.edu.sg}; \url{ypke@ntu.edu.sg}).}
\thanks{Kai He is with the Saw Swee Hock School of Public Health at National University of Singapore, Singapore (e-mail:\url{ kai_he@nus.edu.sg}}
\thanks{Wei Zhang is with the Cognitive Neuroimaging Centre and Lee Kong Chian School of Medicine, Nanyang Technological University, Singapore, 636921 Singapore (e-mail: \url{wilson.zhangwei@ntu.edu.sg}).}
\thanks{Corresponding author: Xia Dong}}

\markboth{Journal of \LaTeX\ Class Files,~Vol.~14, No.~8, August~2021}%
{Shell \MakeLowercase{\textit{et al.}}: A Sample Article Using IEEEtran.cls for IEEE Journals}


\maketitle

\begin{abstract}
Brain network analysis plays a crucial role in identifying distinctive patterns associated with neurological disorders. Functional magnetic resonance imaging (fMRI) enables the construction of brain networks by analyzing correlations in blood-oxygen-level-dependent (BOLD) signals across different brain regions, known as regions of interest (ROIs). These networks are typically constructed using atlases that parcellate the brain based on various hypotheses of functional and anatomical divisions. However, there is no standard atlas for brain network classification, leading to limitations in detecting abnormalities in disorders. Recent methods leveraging multiple atlases fail to ensure consistency across atlases and lack effective ROI-level information exchange, limiting their efficacy. To address these challenges, we propose the Atlas-Integrated Distillation and Fusion network (AIDFusion), a novel framework designed to enhance brain network classification using fMRI data. AIDFusion introduces a disentangle Transformer to filter out inconsistent atlas-specific information and distill meaningful cross-atlas connections. Additionally, it enforces subject- and population-level consistency constraints to improve cross-atlas coherence. To further enhance feature integration, AIDFusion incorporates an inter-atlas message-passing mechanism that facilitates the fusion of complementary information across brain regions. We evaluate AIDFusion on four resting-state fMRI datasets encompassing different neurological disorders. Experimental results demonstrate its superior classification performance and computational efficiency compared to state-of-the-art methods. Furthermore, a case study highlights AIDFusion's ability to extract interpretable patterns that align with established neuroscience findings, reinforcing its potential as a robust tool for multi-atlas brain network analysis. The code is publicly available at \url{https://github.com/AngusMonroe/AIDFusion}.
\end{abstract}


\section{Introduction}
\label{sec:intro}

In the field of neuroscience, a key objective is to identify distinctive patterns associated with neurological disorders (e.g., Alzheimer’s, Parkinson’s, and Autism) by the brain networks~\cite{zhou2012predicting,rudie2013altered}. Resting-state functional magnetic resonance imaging (fMRI) is widely employed among various neuroimaging techniques to characterize the connectivities among brain regions~\cite{biswal1995functional}. 
This results in brain networks where each node represents a specific brain region, referred to as a region of interest (ROI). Each edge indicates a pairwise correlation between the blood-oxygen-level-dependent (BOLD) signals of two ROIs~\cite{smith2011network}, revealing the connectivity between brain regions and indicating which areas tend to be activated synchronously or exhibit correlated activities.

Brain networks model neurological systems as graphs, allowing the use of graph-based techniques to understand their roles and interactions~\cite{kawahara2017brainnetcnn,lanciano2020explainable,wang2023effective}. Constructing these brain networks involves using a specific atlas to parcellate the brain into ROIs. Various atlases based on different hypotheses of brain parcellation, such as anatomical and functional divisions, have been proposed to group similar fMRI regions and create ROIs~\cite{tzourio2002automated,makris2006decreased,Schaefer135632}. Although proper brain parcellation is essential for detecting abnormalities in neurodegenerative disorders~\cite{long2022multi}, there is no golden standard atlas for brain network classification. Relying on a single atlas for brain network analysis has two main drawbacks. First, some voxels may not be assigned to any specific ROI, potentially leading to the loss of important information. Second, each atlas is based on a different parcellation hypothesis. The BOLD signal of an ROI is averaged from all voxels within it, possibly missing detailed information. To address these limitations, recent works have proposed using multiple atlases with different parcellation modes to enhance multi-atlas brain network analysis. Some methods~\cite{chu2022multi,mahler2023pretraining} independently encode brain networks from various atlases and then aggregate the graph representations as a late feature fusion scheme for the final prediction. Another approach \cite{lee2024spectral} incorporates early feature fusion by incorporating multi-atlas information from the raw data and using the fused feature for representation learning. However, these methods (1) neglect the need of consistency across atlases, potentially leading to the under-utilization of cross-atlas information; and (2) lack ROI-level information exchange throughout the entire representation learning process, which could hinder the models' ability to discern complementary information across different atlases.

In this paper, we propose an Atlas-Integrated Distillation and Fusion network (AIDFusion) to address the aforementioned limitations by utilizing atlas-consistent information distillation and cross-atlas complementary information fusion. Specifically, AIDFusion introduces a disentangle Transformer to filter out inconsistent atlas-specific information and distill distinguishable connections across different atlases. Subject- and population-level consistency constraints are applied to enhance cross-atlas consistency. Furthermore, to facilitate the fusion of complementary information across ROIs in multi-atlas brain networks, AIDFusion employs an inter-atlas message-passing mechanism that leverages spatial information. In summary, our key contributions are:

Unlike existing multi-atlas methods that typically apply early or late fusion heuristics, AIDFusion introduces an end-to-end architecture capable of learning consistent and complementary representations from heterogeneous atlases. It achieves this with fewer parameters, faster convergence, and better generalization performance, particularly in out-of-distribution (OOD) scenarios such as cross-site evaluations.

\begin{itemize}
\item \revise{\textbf{A novel architecture for multi-atlas brain network classification.} AIDFusion integrates disentangled representation learning, inter-atlas message passing, and multi-level consistency constraints to effectively leverage complementary information from multiple brain atlases.}
\item \revise{\textbf{Comprehensive evaluation across four resting-state fMRI datasets.} AIDFusion consistently outperforms 11 strong baselines, including both classical ML, general-purposed GNN and  brain network modes, not only in accuracy but also in training efficiency.}
\item \revise{\textbf{Interpretability and robustness.} AIDFusion extracts highly interpretable connectivity patterns that align with known neuroscience findings and demonstrates strong generalization in cross-site experiments, highlighting its clinical applicability and robustness to domain shifts.}
\end{itemize}

\section{Related Work}
\label{sec:related_work}

\subsection{Brain Network Analysis with Various Atlases}

\textbf{Multi-atlas methods} introduce multiple brain atlases for each neuroimage, which can provide information that complements each other and offers ample details without being restricted by the parcellation mode. MGRL~\cite{chu2022multi} pioneered the construction of multi-atlas brain networks using various atlases. It applied graph convolutional networks (GCNs) to learn multi-atlas representations and perform graph-level fusion for disease classification. METAFormer~\cite{mahler2023pretraining} proposed a multi-atlas enhanced transformer approach with self-supervised pre-training for Autism spectrum disorder (ASD) classification. A graph-level late fusion was utilized to aggregate the representations of different atlases. Lee et al. \cite{lee2024spectral} employed a multi-atlas fusion approach that integrates early fusion on the raw feature to capture complex brain network patterns. STW-MHGCN \cite{liu2023deep} constructs a spatial and temporal weighted hyper-connectivity network to fuse multi-atlas information, and Huang et al. \cite{huang2020self} adopt a voting strategy to integrate the classification results of different classifiers (each corresponding to a different atlas) for ASD diagnosis. \jiaxing{CcSi-MHAHGEL~\cite{wang2024multiview} introduces a class-consistency and site-independence Multiview Hyperedge-Aware HyperGraph Embedding Learning framework to integrate brain networks constructed on multiple atlases in a multisite fMRI study.} However, these studies did not consider the inherent consistency between atlases. Independently encoding multi-atlas brain networks without constraints might extract atlas-specific information, distracting from disease-related pattern modeling. Moreover, existing works only incorporate primitive early or late feature fusion between atlases. This absence of intermediate ROI-level interaction could hinder their models' ability to discern complementary information in each atlas. To the best of our knowledge, our work is the first to introduce information distillation with consistency constraints and employ intermediate ROI-level interaction for complementary information fusion. Note that in our work, multiple atlases are applied to preprocessed images for parcellation, meaning our method is based on a single template. 

\textbf{Multi-modal and multi-resolution methods} also explore brain networks using various atlases. Research about multi-modal brain networks~\cite{zhou2019latent,zhou2020multi,zhu2022multimodal,zhang2023multi,qu2024integrated,zhu2024spatio} employed multiple modalities of neuroimaging data, including fMRI, Diffusion Tensor Imaging (DTI) and Positron Emission Tomography (PET), with various atlases to enhance brain network classification, as different modalities provide abundant information compared to a single modality. 
However, these multi-modal methods focus on fusing structural and functional connectivity information instead of trying to capture the whole picture of the single modality data.
Another line of research~\cite{zhang2020multiview,liu2021building,liu2023hierarchical,wen2024multi} focuses on applying multi-resolution atlases to fMRI data to capture individual behavior across coarse-to-fine scales. However, the technical design of these approaches focuses on extracting information from both fine and coarse scales under the same parcellation mode. Although multi-modal and multi-resolution methods employ various atlases, they focus on different objectives from multi-atlas approaches, and the field of brain network analysis with multi-atlas is still in its infancy stage.

\subsection{Graph Neural Networks (GNNs)}

In recent years, Graph Neural Networks (GNNs) have gained significant attention for brain network analysis. Ktena et al. \cite{ktena2017distance} applied graph convolutional networks to learn similarities between pairs of brain networks (subjects). BrainNetCNN \cite{kawahara2017brainnetcnn} introduced edge-to-edge, edge-to-node, and node-to-graph convolutional filters to capture the spatial information within brain networks. MG2G \cite{xu2021graph} employs a two-stage approach, where node representations are initially learned using an unsupervised stochastic graph embedding model based on latent distributions. These representations are then used to train a classifier for identifying significant ROIs associated with Alzheimer's disease (AD). Zhang et al. \cite{zhang2022classification} combined local ROI-GNN and global subject-GNN models to learn the brain network representations. ContrastPool \cite{xu2024contrastive} introduced a dual-attention mechanism to extract discriminative features across ROIs for subjects within the same group.

An alternative approach for graph representation learning involves Transformer-based models \cite{vaswani2017attention}, which adapt the attention mechanism to incorporate global information for each node and positional encoding to capture graph topology. Graph Transformers have gained significant attention due to their impressive performance in graph representation learning \cite{dwivedi2020generalization, ying2021transformers, rampavsek2022recipe}. Several Transformer-based methods have emerged for brain network analysis. For example, one method \cite{kan2022brain} applied Transformers to learn pairwise connection strengths among brain regions across individuals. Com-BrainTF \cite{bannadabhavi2023community} uses a hierarchical local-global transformer to generate community-aware node embeddings, while GBT \cite{peng2024gbt} employs an attention weight matrix approximation to focus on the most relevant components for enhanced graph representation. THC \cite{dai2023transformer} introduced an interpretable Transformer-based model for joint hierarchical cluster identification and brain network classification. DART \cite{kan2023dynamic} segmented BOLD signals to generate dynamic brain networks, which were then integrated with static networks for improved representation learning. Contrasformer \cite{xu2024contrasformer} introduces a contrast graph to highlight the difference between groups and thus improve the model’s generalization ability. Most GNN- and Transformer-based methods for brain network analysis are designed for single-atlas use, which may create a dependency on specific parcellation schemes.

\begin{figure*}[t]
\begin{center}
\includegraphics[width=\linewidth]{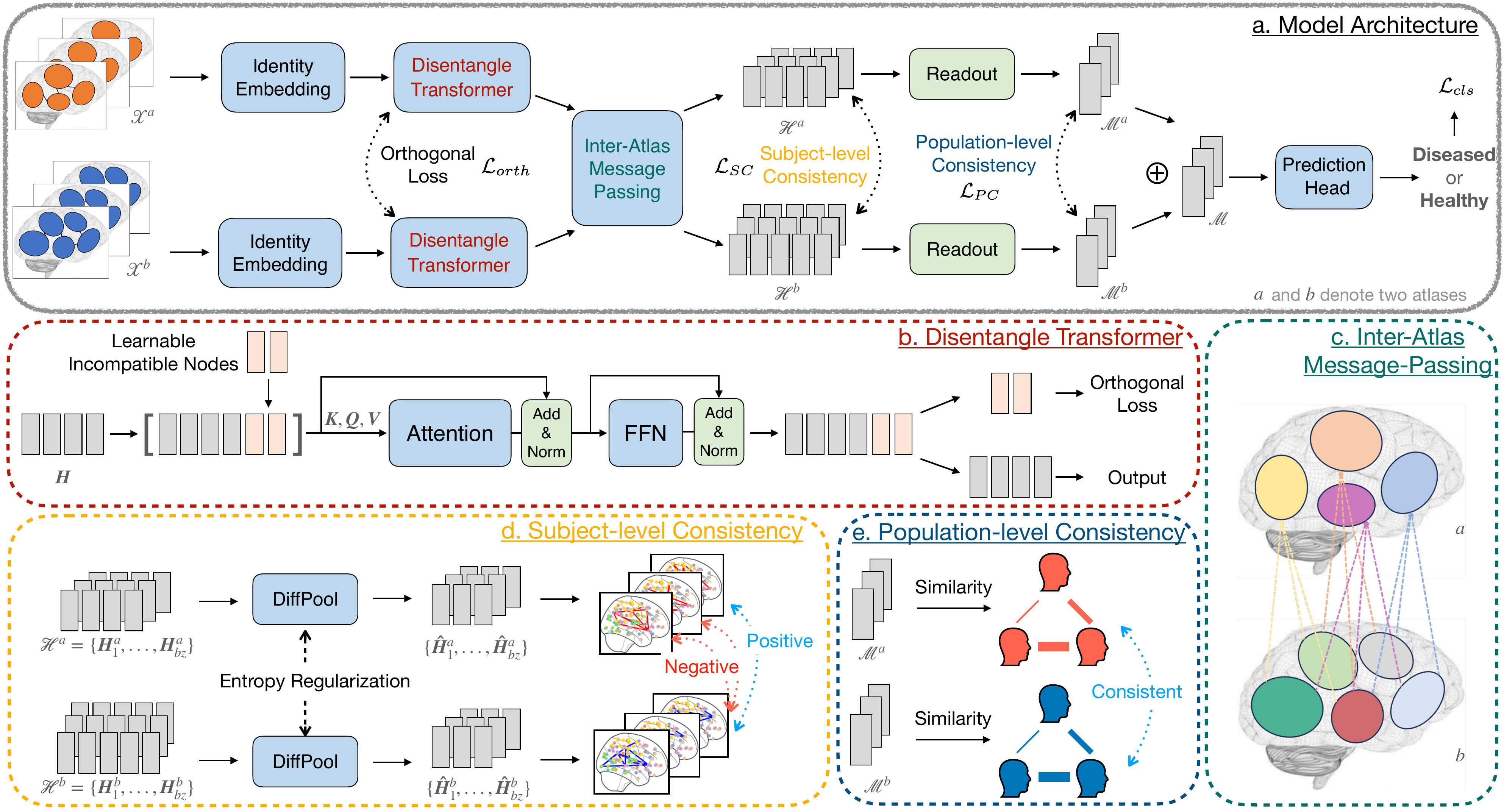}
\end{center}
\caption{The framework of AIDFusion for multi-atlas brain network classification. The proposed framework includes three key components: Disentangle Transformer, Inter-Atlas Message-Passing, and Subject- and Population-level Consistency Constraint.}
\label{fig:framework}
\end{figure*}

\section{Preliminaries}
\label{sec:preliminaries}

\subsection{Brain Network Construction}

We introduce our method using two atlases, $a$ and $b$, for simplicity. It can easily generalize to more atlases. In this work, we use the datasets preprocessed and released by Xu et al. \cite{xu2023data}. Each subject is characterized by two brain networks, represented by connectivity matrices $\boldsymbol{X}^a \in \mathbb{R}^{n_a \times n_a}$ and $\boldsymbol{X}^b \in \mathbb{R}^{n_b \times n_b}$. These matrices are derived using different atlases, which divide the whole brain into $n_a$ and $n_b$ ROIs respectively. In these matrices, each entry represent an edge in the brain network, which is calculated by Pearson's correlations between the region-averaged BOLD signals from pairs of ROIs. Thus these brain networks are weighed and fully-connected. Essentially, these brain networks capture functional relationships between different ROIs.

\subsection{Problem Definition}

Multi-atlas brain network classification aims to predict the distinct class of each subject by using brain networks constructed with various atlases for the same fMRI data. Given a labeled brain network dataset $\mathcal{D} = \{(\boldsymbol{X}^a, \boldsymbol{X}^b, y_X)\}$, where $y_X$ is the class label of brain networks $\boldsymbol{X}^a$ and $\boldsymbol{X}^b$, the objective of multi-atlas brain network classification is to learn a predictive function $f$: $(\boldsymbol{X}^a, \boldsymbol{X}^b) \rightarrow y_X$ that can map the input brain networks to the groups they belong to, expecting that the function $f$ also works well on the other unseen brain networks.

\section{Methodology}
\label{sec:method}

In this section, we provide a detailed exposition of the design of our proposed Atlas-Integrated Distillation and Fusion network (AIDFusion), depicted in Figure \ref{fig:framework}. Two brain networks constructed with different atlases are separately processed in our model.
In the following, we first introduce the disentangle Transformer with identity embedding to remove inconsistent atlas-specific information (Section \ref{subsec:disen_tf}). We then describe the inter-atlas message-passing for spatial-based intense fusion of cross-atlas (Section \ref{subsec:ia_mp}). Finally, we discuss our design of the losses that enforce atlas-consistent information distillation with domain considerations (Section \ref{subsec:loss}).

\subsection{Disentangle Transformer with Identity Embedding}
\label{subsec:disen_tf}

\noindent
\textbf{Identity Embedding.} In graph Transformer models, positional embedding is commonly used to encode the topological information of the graph. However, designs like distance-based, centrality-based, and eigenvector-based positional embeddings~\cite{li2020distance,ying2021transformers,wang2022equivariant} are impractical for brain networks due to their high density (always fully connected). Correlation-based brain networks already contain sufficient positional information for ROIs, making general positional embeddings both costly and redundant. Instead, we propose a learnable identity embedding that adaptively learns a unique identity for each ROI, aligning nodes in different subjects that correspond to the same ROI in the same atlas. This embedding assigns the same identity to nodes within the same ROI. The design is inspired by the concept of word embeddings in natural language processing~\cite{mikolov2013efficient}, where each word is assigned an independent learnable representation.
As shown in Eq. (\ref{eq:id_enc}), we introduce a parameter matrix $\boldsymbol{W}_{ID}$ to encode node identities alongside original node features $\boldsymbol{X}$, with $\operatorname{MLP}(\cdot)$ denoting a multilayer perceptron (MLP).
\begin{equation}
\label{eq:id_enc}
\small
\boldsymbol{H}_{ID} = \boldsymbol{X} + \operatorname{MLP}(\boldsymbol{X} + \boldsymbol{W}_{ID}).
\end{equation}
\textbf{Disentangle Transformer.} Introducing learnable tokens in the input sequence of a Transformer has been a method used to capture global information. In natural language processing, Burtsev et al. \cite{burtsev2020memory} first utilized a learnable $[CLS]$ token to improve machine translation tasks. In computer vision, Darcet et al. \cite{darcet2023vision} introduced register tokens to avoid recycling tokens from low-informative areas. Motivated by these prior works, we propose a disentangle Transformer to filter out inconsistent atlas-specific information by introducing incompatible nodes. We elaborate this module in Figure \ref{fig:framework}b. Specifically, given an identity-encoded graph feature matrix $\boldsymbol{H}_{ID} \in \mathbb{R}^{n \times d}$, where $n$ is the number of nodes and $d$ is the hidden dimension, we add $r$ learnable incompatible nodes $\boldsymbol{W}_{INC} \in \mathbb{R}^{r \times d}$ to the feature matrix:
\begin{equation}
\label{eq:redundant}
\small
\boldsymbol{H}^{\prime}=\left [
        \begin{array}{c}        
        \boldsymbol{H}_{ID}\\
        \boldsymbol{W}_{INC}\\
        \end{array}
        \right ], 
\end{equation}
where $\left [ \begin{array}{c}\cdot\\ \cdot\\ \end{array} \right ]$ denotes the append operation. To enforce each incompatible node captures different information, we initialize them using the Gram-Schmidt process~\cite{cheney2009linear} to ensure they are orthogonal to each other. Then the self-attention function~\cite{vaswani2017attention} is applied to $\boldsymbol{H}^{\prime} \in \mathbb{R}^{(n+r) \times d}$:
\begin{equation}
\label{eq:attn}
\small
\operatorname{Attn}(\boldsymbol{H}^{\prime}) = \operatorname{norm} \left(\boldsymbol{H}^{\prime} + \operatorname{softmax}\left(\frac{\boldsymbol{Q} \boldsymbol{K}^\mathsf{T}}{\sqrt{n+r}}\right) \boldsymbol{V}\right),
\end{equation}
\begin{equation}
\label{eq:qkv}
\small
\boldsymbol{Q}=\boldsymbol{H}^{\prime}\boldsymbol{W}_Q, \boldsymbol{K}=\boldsymbol{H}^{\prime}\boldsymbol{W}_K, \boldsymbol{V}=\boldsymbol{H}^{\prime}\boldsymbol{W}_V,
\end{equation}
%
%
\noindent
where $\boldsymbol{W}_Q, \boldsymbol{W}_K, \boldsymbol{W}_V \in \mathbb{R}^{d \times d}$ are parameter matrices and $\operatorname{norm}(\cdot)$ is a layer normalization. 

In addition to the attention layer, a position-wise feed-forward network (FFN) with a layer normalization function is applied to each position to get the output node representations. The brain network of each atlas goes through a separate Disentangle Transformer. At the output of the disentangle Transformer, the incompatible nodes are discarded and only the ROI nodes are used.

\noindent
\textbf{Orthogonal Loss.} As the brain networks derived from different atlases are based on the same fMRI data, we aim to ensure they contain similar information by filtering out inconsistent atlas-specific information. Therefore, we propose an orthogonal loss to enforce the representations of incompatible nodes to be orthogonal to each other across all atlases by minimizing their dot product:
\begin{equation}
\small
\label{eq:orth_loss}
\mathcal{L}_{orth} = \frac{1}{r} \sum \frac{||\boldsymbol{W}_{INC}^a \cdot \boldsymbol{W}_{INC}^b||}{||\boldsymbol{W}_{INC}^a|| \cdot ||\boldsymbol{W}_{INC}^b||}.
\end{equation}

\subsection{Inter-Atlas Message-Passing}
\label{subsec:ia_mp}

The features in different atlases originate from totally different parcellation modes. Pulling those highly correlated features of two different atlases into a shared space allows their effective fusion. Existing literature on multi-atlas brain networks independently learns the representations of ROIs in each atlas without exchanging information across atlases~\cite{chu2022multi, lee2024spectral}. Additionally, the spatial relationship between ROIs in different atlases is neglected in these works. Our proposed AIDFusion enables inter-atlas message-passing between neighboring regions in different atlases by considering spatial information.  Specifically, for each ROI, we define its spatial coordinate as the centroid of the voxels assigned to that region and compute the Euclidean distance between ROIs in different atlases as the spatial distance.
As shown in Figure \ref{fig:framework}c, we utilize the $k$-nearest-neighbor ($k$NN) algorithm to connect each ROI to $k$ ROIs from the other atlas. Note that we only construct inter-atlas connections without considering intra-atlas connections since the information exchange within the same atlas has already been processed in the previous disentangle Transformer. 
When atlases are derived from different templates (e.g., MNI vs. Talairach), coordinate alignment becomes a critical issue. Fortunately, spatial normalization and template conversion between standard spaces is a well-established task in neuroimaging. Several mature techniques~\cite{crum2004non,lancaster2007bias,lacadie2008more,laird2010comparison} exist to handle such transformations accurately. These methods can be readily incorporated into our framework as a preprocessing step to ensure spatial compatibility before applying inter-atlas message-passing.

Afterwards, an adjacency matrix $\boldsymbol{A}^{ab} \in \{0, 1\}^{(n^a + n^b) \times (n^a + n^b)}$ is obtained and used for graph convolution~\cite{kipf2016semi}:
\begin{equation}
\label{eq:ia_mp}
\small
\operatorname{GCN}(\boldsymbol{A}^{ab}, \boldsymbol{H}^{ab}) = \sigma\left(\boldsymbol{D}^{-\frac{1}{2}}\boldsymbol{A}^{ab}\boldsymbol{D}^{-\frac{1}{2}}\boldsymbol{H}^{ab}\boldsymbol{W}_{GC}\right),
\end{equation}
where $\sigma$ is the activation function (e.g., ReLU), $\boldsymbol{D}$ is the degree matrix of $\boldsymbol{A}^{ab}$, $\boldsymbol{H}^{ab} \in \mathbb{R}^{(n^a + n^b) \times d}$ is the combined node representation matrix for the two atlases, and $\boldsymbol{W}_{GC}$ is the learnable weight matrix of the GCN layer. 

\subsection{Subject- and Population-level Consistency}
\label{subsec:loss}

\noindent
\textbf{Subject-level Consistency.} To ensure the high-level consistency for the two brain networks from different atlases, we introduce a contrastive loss on the subject level. First, we apply DiffPool~\cite{ying2018hierarchical} to each atlas to capture higher-level patterns. The DiffPool contains two GCN layers. $\operatorname{GCN}_{pool}$ is used to learn a cluster assignment matrix $\boldsymbol{S} \in \mathbb{R}^{n \times n^{\prime}}$ as shown in Eq. (\ref{eq:assign_matrix}). Herein, $n^{\prime}$ denotes the number of clusters, which is controlled by a preset pooling ratio. The other $\operatorname{GCN}_{emb}$ is used to generate the embedded feature $\boldsymbol{Z} \in \mathbb{R}^{n \times d}$ as shown in Eq. (\ref{eq:emb_gcn}). Both these two GCNs are defined similarly with Eq. (\ref{eq:ia_mp}). We sparsify the connectivity matrices $\boldsymbol{X}$ by keeping top 20\% correlations and use them as the adjacency matrices $\boldsymbol{A}$ in these two GCNs, to avoid over-smoothing. The feature matrices of two atlas $\boldsymbol{H}$ are obtained from the output of inter-atlas message-passing.
\begin{equation}
\label{eq:assign_matrix}
\small
\boldsymbol{S} = \operatorname{softmax}\left( \operatorname{GCN}_{pool}(\boldsymbol{A}, \boldsymbol{H}) \right).
\end{equation}
\begin{equation}
\label{eq:emb_gcn}
\small
\boldsymbol{Z}= \operatorname{GCN}_{emb}(\boldsymbol{A}, \boldsymbol{H}).
\end{equation}
\begin{table*}[h]
\centering
\caption{\jiaxing{Statistics and class information of brain network datasets used in this work.}}
\begin{tabular}{cccccc}
\hline
Dataset & Condition        & Sex (F/M) & Age (mean ± std)  & Subject\# & Class\#  \\ \hline
ABIDE   & Autism Spectrum   Disorder       & 152/873   & 16.5 ± 7.4  & 1025      & 2     \\
ADNI    & Alzheimer’s   Disease        & 728/599 & 74.6 ± 7.9        & 1326      & 6     \\
PPMI    & Parkinson’s   Disease      & 82/127 & 62.9 ± 9.5          & 209       & 4   \\
Matai   & Mild Traumatic   Brain Injury (mTBI)  & N/A & N/A & 60        & 2   \\ \hline
\end{tabular}
\label{tab:dataset_statistic}
\end{table*}
A new feature matrix $\hat{\boldsymbol{H}} \in \mathbb{R}^{n^\prime \times d}$ is generated by the cluster assignment matrix $\boldsymbol{S}$ and the embedded node feature matrix $\boldsymbol{Z}$ as $\hat{\boldsymbol{H}}=\boldsymbol{S}^\mathsf{T} \boldsymbol{Z}$. This coarsening process aims to generate super-nodes to get higher-level node representations. To avoid GNN treating each ROI and each node cluster equally, we adopt an entropy regularization to the assignment matrices of each atlas:
\begin{equation}
\small
\mathcal{L}_{E}= \frac{1}{n^{\prime}} \sum^{n^{\prime}}_{\jiaxing{j}=1} \left( \operatorname{entropy} ( \boldsymbol{S}^a[\jiaxing{j},:] ) + \operatorname{entropy} ( \boldsymbol{S}^b[\jiaxing{j},:] ) \right), 
\label{eq:entropy_loss}
\end{equation}
\begin{equation}
\small
\operatorname{entropy}(\boldsymbol{p}) = - \sum_{j=1}^{n^\prime} \boldsymbol{p}_j \log(\boldsymbol{p}_j).
\end{equation}
We elaborate on the module of subject-level consistency in Figure \ref{fig:framework}d. Through two DiffPool layers, we align the number of nodes and produce high-quality representations for each atlas. Then we are able to apply a contrastive loss to them by considering representations from the same subject as positive pairs $\mathcal{P}^{pos} = \{ (\hat{\boldsymbol{H}}^a_i, \hat{\boldsymbol{H}}^b_i) \}: i = 1, \dots, bz\}$ and representations from different subjects as negative pairs $\mathcal{P}^{neg} = \{ (\hat{\boldsymbol{H}}^a_i, \hat{\boldsymbol{H}}^b_{\neg i}): i = 1, \dots, bz\}$:
\begin{equation}
\label{eq:sc_loss}
\small
\mathcal{L}_{SC} = - \operatorname{log} \frac {\sum \sum_{(\hat{\boldsymbol{H}}^a_i, \hat{\boldsymbol{H}}^b_i) \in \mathcal{P}^{pos}} \operatorname{exp}(\operatorname{sim}(\hat{\boldsymbol{H}}^a_i, \hat{\boldsymbol{H}}^b_i) / \tau)} {\sum \sum_{(\hat{\boldsymbol{H}}^a_i, \hat{\boldsymbol{H}}^b_{\neg i}) \in \mathcal{P}^{neg}} \operatorname{exp} (\operatorname{sim}(\hat{\boldsymbol{H}}^a_i, \hat{\boldsymbol{H}}^b_{\neg i}) / \tau)},
\end{equation}
where $\tau$ is a temperature hyper-parameter to control the smoothness of the probability distribution~\cite{you2020graph}, $bz$ is the batch size, and $\operatorname{sim}(\cdot)$ denotes the cosine similarity function that is applied to the same row in the two matrices.

\noindent
\textbf{Population-level Consistency.} The readout function $\boldsymbol{m}=\operatorname{READOUT}\left(\boldsymbol{H}\right)$ is an essential component of learning the graph-level representations $\boldsymbol{m} \in \mathbb{R}^{d}$ for brain network analysis (e.g., classification), which maps a set of learned node-level embeddings to a graph-level embedding. 
%
%
To further constrain the consistency for graph representations across different atlases, we introduce a mean squared error (MSE) loss on the population level. Different from some existing works, such as FTMMR~\cite{son2024ftmmr}, that directly enforce the alignment of different views on graph level, we focus on maintaining the relative similarity structure among subjects across atlases.
As shown in Figure \ref{fig:framework}e, a population graph $\boldsymbol{G}$ is constructed by computing the similarity of each two subjects' graph representations in the same atlas. The intuition here is we aim to maintain the relationship of subjects across atlases, instead of directly enforcing graph representations of two atlases to be the same. This constraint does not rely on the target label, as we only encourage the consistency of (dis)similarity across atlases rather than enforcing the similarity within each labeled group. Such loss is formulated as follows:
\begin{equation}
\label{eq:pc_loss}
\small
\mathcal{L}_{PC} = \frac{1}{bz} \sum (\boldsymbol{G}^a - \boldsymbol{G}^b)^2,
\boldsymbol{G}[i,j] = \operatorname{sim}(\boldsymbol{m}_i, \boldsymbol{m}_j), \boldsymbol{m}_i, \boldsymbol{m}_j \in \mathcal{M},
\end{equation}
where $\mathcal{M}$ is the set of graph representations in a batch.

\noindent
\textbf{Total Loss.} The model is supervised by a widely-used cross-entropy loss $\mathcal{L}_{cls}$~\cite{cox1958regression} for brain network classification. The total loss is computed by:
\begin{equation}
\label{eq:total_loss}
\small
\mathcal{L}_{total} = \mathcal{L}_{cls} + \lambda_1 * \mathcal{L}_{SC} + \lambda_2 * \mathcal{L}_{PC} + \lambda_3 * \mathcal{L}_{E} + \lambda_4 * \mathcal{L}_{orth},
\end{equation} 
where $\lambda_1$, $\lambda_2$, $\lambda_3$ and $\lambda_4$ are trade-off hyperparameters for balancing different losses.

\section{Experimental Results}
\label{sec:exp}


\subsection{Brain Network Datasets}  

We use four brain network datasets released by Xu et al.~\cite{xu2023data} to study various disorders: ABIDE~\cite{craddock2013neuro} for Autism Spectrum Disorder (ASD), ADNI~\cite{dadi2019benchmarking} for Alzheimer's Disease (AD), PPMI~\cite{badea2017exploring} for Parkinson's Disease (PD), and Mātai for mild traumatic brain injury (mTBI)~\cite{xu2023data}. A summary of the brain network datasets is provided in Table \ref{tab:dataset_statistic}. The atlases used in this study are Schaefer~\cite{Schaefer135632} and AAL~\cite{tzourio2002automated}, with 100 and 116 ROIs, respectively. We choose these two atlases because they are among the most widely used and well-established brain parcellation schemes in the neuroimaging community. Importantly, they represent two complementary perspectives: the AAL atlas is based on anatomical landmarks, while the Schaefer atlas is derived from functional connectivity patterns.

\noindent
\textbf{ABIDE} The ABIDE initiative aggregates functional brain imaging data from multiple research labs worldwide to support ASD research. The age range of participants spans from 5 to 64 years. The dataset classifies subjects into two groups: typical controls (TC) and individuals diagnosed with ASD.

\noindent
\textbf{ADNI} The raw images used in this study were obtained from the ADNI database (\url{adni.loni.usc.edu}). Launched in 2003 as a public-private partnership led by Principal Investigator Michael W. Weiner, MD, the primary goal of ADNI is to test whether serial MRI, PET, other biological markers, and clinical assessments can be combined to measure the progression of mild cognitive impairment (MCI) and early Alzheimer's Disease (AD). For up-to-date information, visit \url{www.adni-info.org}. We include subjects from six distinct stages of AD: cognitively normal (CN), significant memory concern (SMC), MCI, early MCI (EMCI), late MCI (LMCI), and AD.

\noindent
\textbf{PPMI} The Parkinson's Progression Markers Initiative (PPMI) is a comprehensive study aimed at identifying biological markers associated with Parkinson's risk, onset, and progression. The PPMI dataset includes multimodal and multi-site MRI images from subjects in four distinct groups: normal controls (NC), scans without evidence of dopaminergic deficit (SWEDD), prodromal Parkinson’s disease, and diagnosed PD.

\noindent
\textbf{Mātai} The Mātai study is a longitudinal, single-site, single-scanner investigation designed to detect subtle brain changes due to a season of contact sports. This dataset consists of brain network data collected from the Gisborne-Tairāwhiti region of New Zealand, with 35 contact sport players imaged pre-season (N=35) and post-season (N=25). Subtle brain changes confirmed by diffusion imaging studies were observed as a result of playing contact sports.

\subsection{Baseline Models}

\jiaxing{We use 8 single-atlas methods and 6 multi-atlas methods as baselines to evaluate our proposed AIDFusion, including: (1) \textbf{Conventional machine learning (ML) models}: Logistic Regression (LR) and Support Vector Machine Classifier (SVM) from scikit-learn~\cite{pedregosa2011scikit}. These models take the flattened upper-triangle connectivity matrix as vector input, instead of using the brain network. (2) \textbf{General-purposed GNNs}: GCN~\cite{kipf2016semi} and Transformer~\cite{vaswani2017attention}. (3) \textbf{Single-Atlas Models tailored for brain networks}: BrainNetCNN~\cite{kawahara2017brainnetcnn}, MG2G~\cite{xu2021graph}, ContrastPool~\cite{xu2024contrastive} and BNT~\cite{kan2022brain}. (4) \textbf{Multi-atlas models}: MultiLR (multi-atlas version of LR, concatenate the flatten feature of multiple atlases as input), MultiSVM (multi-atlas version of SVM, similar with MultiLR); MGRL~\cite{chu2022multi}; MGT (a multi-atlas version of Transformer with the same fusion mechanism as MGRL), METAFormer~\cite{mahler2023pretraining} and LeeNet~\cite{lee2024spectral}.}

The experimental settings are based on those outlined in Xu et al.~\cite{xu2024contrastive}. For each dataset, we use an 8:1:1 split for training, validation, and testing, respectively. Each model is evaluated across 5 random seeds using 10-fold cross-validation (CV), with the average accuracy reported for the main results. 
We apply an early stopping criterion: the learning rate is halved if there is no improvement in validation loss over 25 epochs, and training stops once the learning rate reaches a pre-defined minimum threshold. All experiments were conducted on a server equipped with an AMD Ryzen Threadripper PRO 5995WX (64 cores) and an NVIDIA GeForce RTX 4090.

All the baselines used in this paper are implemented by ourselves. In AIDFusion, we adopt a mean pooling layer as the readout function and a two-layer MLP with ReLU as the prediction head. The number of incompatible nodes $r$ is set to 4. The number of clusters $n^\prime$ for the cluster assignment matrix $\boldsymbol{S}$ is set to half of the average number of nodes for all atlases, e.g., for Schaefer100 and AAL116, $n^\prime = (100 + 116) / 2 / 2 = 54$. The temperature hyper-parameter $\tau$ in Eq. (\ref{eq:sc_loss}) is set to 0.75. The hidden dimensions of all layers are set to 100. The batch size of each dataset is set to 10\% of the subject number in the dataset. We tuned the other hyperparameters on validation set including $k$ in $k$NN for the inter-atlas message-passing, trade-off hyperparameters $\lambda_1, \lambda_2, \lambda_3, \lambda_4$ for loss function Eq. (\ref{eq:total_loss}), the initial learning rate $init\_lr$, and the minimum learning rate $min\_lr$ used for early stop. To be specific, we search $k$ from \{3, 5, 10\}, $\lambda_1$ from \{1e-2, 1e-1, 1e0, 1e1, 1e2\}, $\lambda_2$ from \{1e-2, 1e-1, 1e0, 1e1, 1e2\}, $\lambda_3$ from \{1e-6, 1e-5, 1e-4, 1e-3, 1e-2\}, $\lambda_4$ from \{1e-1, 1e0, 1e1\}, $inir\_lr$ from \{5e-5, 8e-5, 1e-4, 2e-4, 7e-4\} and $inir\_lr$ from \{6e-5, 1e-5, 5e-6, 1e-6\}. 
The optimized hyperparameters for AIDFusion are reported in Table~\ref{tab:optimized_params}.

\begin{table}[h]
\caption{The optimized hyperparameters for AIDFusion.}
\centering
\begin{tabular}{ccccc}
\hline
             & ABIDE & ADNI & PPMI & Mātai \\ \hline
$k$            & 10    & 5    & 5    & 5     \\
$\lambda_1$    & 1e1   & 1e1 & 1e1  & 1e-2  \\
$\lambda_2$    & 1e-1  & 1e1  & 1e1  & 1e0  \\
$\lambda_3$    & 1e-2  & 1e-5 & 1e-4 & 1e-3  \\
$\lambda_4$    & 1e0   & 1e0  & 1e0  & 1e0  \\
$init\_lr$     & 1e-4  & 8e-5 & 1e-4 & 7e-4  \\
$min\_lr$      & 6e-5  & 6e-5 & 6e-5 & 6e-5  \\ \hline
\end{tabular}
\label{tab:optimized_params}
\end{table}

\jiaxing{For machine learning baselines (LR, SVM), we use the official code
of Xu et al. \cite{xu2023data} and implement the multi-atlas version on the top of it. We follow the same parameter-tuning scheme as they do. The full list of parameters, including those we used for the grid search, is provided below. The parameter tuning for MultiLR and MultiSVM are the same with LR and SVM, respectively.}

\begin{table*}[h]
\centering
\caption{Brain Network Classification Results (Average Accuracy ± Standard Deviation) over 5 runs. The first and second best results on each dataset are highlighted in \textbf{bold} and \underline{underline}. \jiaxing{The p-values of one-sided paired t-tests comparing our AIDFusion with the best baselines on the four datasets are 0.0483, 0.0040, 0.0141, and 0.0068, respectively.}}
\begin{tabular}{cccccc}
\hline
atlas & model        & ABIDE    & ADNI           & PPMI           & Mātai           \\ \hline
\multirow{8}{*}{Schaefer100}            & LR       & 64.81   ± 0.00 & 61.97   ± 0.00 & 56.48   ± 0.00 & 60.00   ± 0.00 \\
      & SVM          & 64.41   ± 0.00 & 61.52   ± 0.00 & 63.21   ± 0.00 & 56.67   ± 0.00 \\
      & GCN          & 60.16 ± 1.76   & 60.62 ± 0.42   & 52.60 ± 1.58   & 60.33 ± 3.80   \\
      & Transformer  & 60.34 ± 1.75   & 63.91 ± 0.82   & 58.67 ± 1.94   & 60.00 ± 3.73   \\
      & BrainNetCNN  & 64.93 ± 0.79   & 63.42 ± 1.21   & 53.53 ± 2.26  & 60.33 ± 7.94   \\
      & MG2G         & 63.51 ± 0.30   & 62.98 ± 0.39   & 56.46 ± 1.45  & 61.67 ± 2.36   \\
      & ContrastPool & \underline{65.40} ± 1.01   & 65.13 ± 0.96   & 63.79 ± 1.88   & 62.00 ± 4.77   \\ 
      & \jiaxing{BNT} & \jiaxing{60.56 ± 1.19} & \jiaxing{66.14 ± 0.65} & \jiaxing{57.31 ± 1.37} & \jiaxing{60.67 ± 2.79} \\ \hline
\multirow{8}{*}{AAL116}                 & LR       & 63.80   ± 0.00 & 64.06   ± 0.00 & 56.00   ± 0.00 & 66.67   ± 0.00 \\
      & SVM          & 65.72   ± 0.00 & 63.40   ± 0.00 & \underline{64.12}   ± 0.00 & 65.00   ± 0.00 \\
      & GCN          & 59.50 ± 0.62   & 61.02 ± 0.75   & 51.20 ± 1.63   & 56.67 ± 4.25   \\
      & Transformer  & 60.05 ± 0.78   & 63.52 ± 0.47   & 60.87 ± 1.88   & 62.67 ± 4.18   \\
      & BrainNetCNN  & 63.94 ± 1.00   & 63.09 ± 0.90   & 50.79 ± 1.04   & 68.33 ± 2.64   \\
      & MG2G         & 63.32 ± 1.40   & 63.34 ± 0.70   & 58.76 ± 1.72   & \underline{69.00} ± 3.03   \\
      & ContrastPool & 64.92 ± 0.90   & 66.18 ± 1.13   & 63.22 ± 1.35   & 64.67 ± 3.21   \\ 
       & \jiaxing{BNT} & \jiaxing{59.65 ± 1.44} & \jiaxing{60.73 ± 0.87} & \jiaxing{55.63 ± 1.76} & \jiaxing{63.00 ± 4.63} \\ \hline
\multirow{7}{*}{\begin{tabular}[c]{@{}c@{}c@{}}Schaefer100 \\ + \\  AAL116\end{tabular}} & MultiLR  & 65.23   ± 0.00 & 64.99   ± 0.00 & 55.00   ± 0.00 & 56.67   ± 0.00 \\
& MultiSVM & 64.31   ± 0.00 & 65.21   ± 0.00 & 63.60   ± 0.00 & 58.33   ± 0.00 \\
& MGRL         & 60.14 ± 0.94   & 62.65 ± 0.90   & 54.37 ± 1.37  & 68.66 ± 3.61   \\
& MGT          & 63.71 ± 1.73   & 63.36 ± 0.49   & 62.62 ± 1.55   & 65.33 ± 3.21   \\
& METAFormer   & 61.97 ± 0.87   & \underline{66.76} ± 0.28   & 54.82 ± 1.91   & 67.64 ± 6.70   \\
& LeeNet       & 61.33 ± 0.54   & 64.60 ± 0.17   & 60.83 ± 0.21   & 58.33 ± 0.0   \\
& AIDFusion (ours)  & \textbf{66.51} ± 1.45   & \textbf{67.78} ± 0.35   & \textbf{66.04} ± 1.28   & \textbf{73.67} ± 1.39   \\ \hline
\end{tabular}
\label{tab:main_results}
\end{table*}

\begin{table}[h]
\caption{Results of more evaluation metrics (precision, recall, micro-F1, and ROC-AUC) on ABIDE dataset over 5 runs. The best result is highlighted in \textbf{bold}. The p-values of one-sided paired t-tests for all metrics except precision are under 0.05.}
\label{tab:other_metrics}
\centering
\scalebox{0.9}{
\begin{tabular}{c|cccc}
\hline
  & Precision    & Recall        & micro-F1     & ROC-AUC         \\ \hline
MGRL       & 59.81 ± 3.12 & 59.79 ± 3.05 & 59.59 ± 2.98 & 61.36 ± 2.01 \\
MGT        & 60.89 ± 3.21 & 67.75 ± 5.68 & 64.08 ± 1.77 & 61.47 ± 1.52 \\
METAFormer & 60.06 ± 0.98 & 61.31 ± 0.61 & 60.52 ± 0.67 & 61.94 ± 0.83 \\
LeeNet     & \textbf{64.50} ± 0.86 & 43.53 ± 0.99 & 51.58 ± 0.99 & 60.52 ± 0.55 \\
AIDFusion  & 62.31 ± 3.02 & \textbf{73.71} ± 6.03 & \textbf{67.69} ± 1.67 & \textbf{66.57} ± 1.15 \\ \hline
\end{tabular}}
\end{table}

\begin{table}[h]
\caption{Results of more atlases on ABIDE dataset over 10-fold-CV in a single run. The best results for each atlas setting are highlighted in \textbf{bold}. The p-value of one-sided paired t-test for the 3-atlas setting is under 0.1, while p-values for the other 2-atlas settings are under 0.05.}
\label{tab:atlas_results_basc}
\centering
\begin{tabular}{ccccc}
\hline
\multicolumn{3}{c}{Atlas}                                    & \multirow{2}{*}{model} & \multirow{2}{*}{acc ± std} \\
Schaefer100        & AAL116             & BASC122               &                        &                            \\ \hline
\multirow{5}{*}{\checkmark} & \multirow{5}{*}{\checkmark} & \multirow{5}{*}{}  & MGRL                   & 61.56 ± 4.90               \\
                   &                    &                    & MGT                    & 63.32 ± 3.90               \\
                   &                    &                    & METAFormer             & 61.27 ± 4.05               \\
                   &                    &                    & LeeNet                 & 61.28 ± 3.12               \\
                   &                    &                    & AIDFusion                   & \textbf{66.35} ± 3.26               \\ \hline
\multirow{5}{*}{}  & \multirow{5}{*}{\checkmark} & \multirow{5}{*}{\checkmark} & MGRL                   & 60.80 ± 5.12               \\
                   &                    &                    & MGT                    &   58.49 ± 5.64             \\
                   &                    &                    & METAFormer             &    62.92 ± 5.79            \\
                   &                    &                    & LeeNet                 &    60.01 ± 4.00            \\
                   &                    &                    & AIDFusion                   & \textbf{65.97} ± 3.60              \\ \hline
\multirow{5}{*}{\checkmark} & \multirow{5}{*}{}  & \multirow{5}{*}{\checkmark} & MGRL                   & 63.14 ± 4.17              \\
                   &                    &                    & MGT                    & 59.90 ± 3.46               \\
                   &                    &                    & METAFormer             & 62.53 ± 3.78               \\
                   &                    &                    & LeeNet                 & 59.12 ± 4.20               \\
                   &                    &                    & AIDFusion                   & \textbf{64.79} ± 2.80               \\ \hline
\multirow{5}{*}{\checkmark} & \multirow{5}{*}{\checkmark} & \multirow{5}{*}{\checkmark} & MGRL                   & 63.62 ± 4.57               \\
                   &                    &                    & MGT                    & 62.84 ± 3.85               \\
                   &                    &                    & METAFormer             & 65.80 ± 5.61               \\
                   &                    &                    & LeeNet                 & 60.19 ± 3.77               \\
                   &                    &                    & AIDFusion                   & \textbf{66.65} ± 4.14               \\ \hline
\end{tabular}
\end{table}

\subsection{Main Results}

The classification accuracy on 4 brain network datasets over 5 runs is reported in Table \ref{tab:main_results}. For certain diseases, the effectiveness/informativeness of different atlases is different. On Matai, 6 out of 7 baselines attain better performance with AAL116 than with Schaefer100. On ADNI, 4 out of 7 baselines also perform better with AAL116. In contrast on ABIDE, 6 out of 7 baselines achieve better results with Schaefer100 than with AAL116. It demonstrates the importance of using multi-atlas for brain network analysis instead of relying on one specific atlas. Moreover, it is evident that the multi-atlas baselines with a simple late fusion mechanism (MGRL and MGT) outperform their respective single-atlas models (GCN and Transformer) in the most cases. This highlights the effectiveness of multi-atlas approaches in enhancing the performance of base models. However, conventional ML models (MultiLR and MultiSVM) fail to outperform their single-atlas versions in more cases, possibly due to their inability to effectively utilize multi-atlas features with simple concatenate fusion.

We can also observe that our proposed AIDFusion consistently outperforms not only all single-atlas methods but also the state-of-the-art multi-atlas methods across all these datasets. Specifically, AIDFusion achieves improvements over all multi-atlas methods on these four datasets by up to 6.77\% ((73.67\% - 69.00\%) / 69.00\% = 6.77\% on Mātai). Our model gains more improvement on small datasets (PPMI and Mātai) than on large datasets (ABIDE and ADNI), which meets the intuition that information utilization tends to be more critical in applications with smaller sample sizes. 

In addition to accuracy, we also report other evaluation metrics, including precision, recall, micro-F1, and ROC-AUC, for all the multi-atlas deep models on the ABIDE dataset. As displayed in Table \ref{tab:other_metrics}, the proposed AIDFusion performs the best across all these evaluation metrics except for precision. We observe that compared to other baselines, our AIDFusion can significantly improve recall without compromising precision. Moreover, in medical diagnostics, it is crucial to ensure that all individuals with a certain condition are correctly identified, even if it leads to some false positives. Missing a true positive (failing to diagnose a disease) can have severe consequences, while false positives can be further examined or retested. Therefore, models with higher recall rates, such as our AIDFusion, are more suitable for real-life medical auxiliary diagnosis.

\begin{figure*}[h]
\centering
\includegraphics[width=0.4\textwidth]{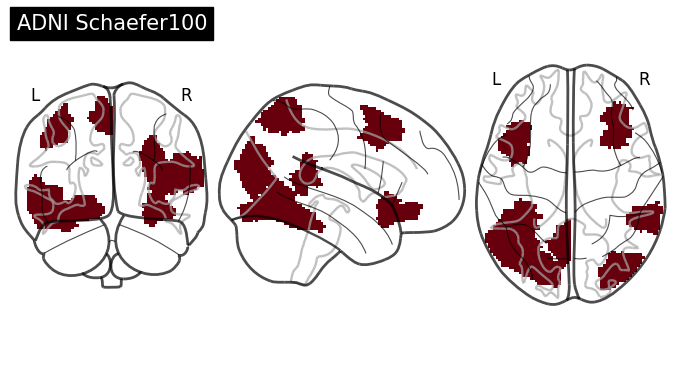}
\includegraphics[width=0.34\textwidth]{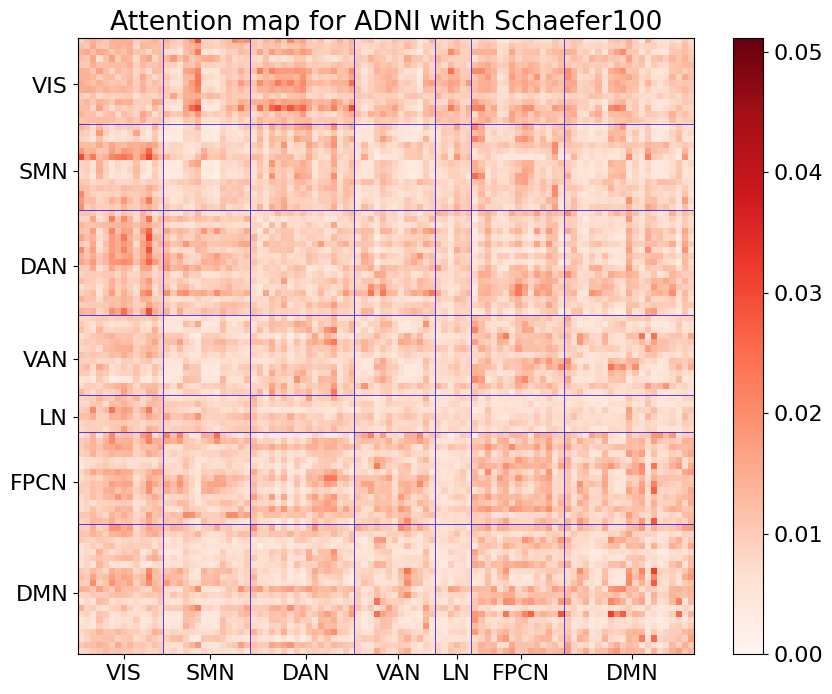}
\includegraphics[width=0.4\textwidth]{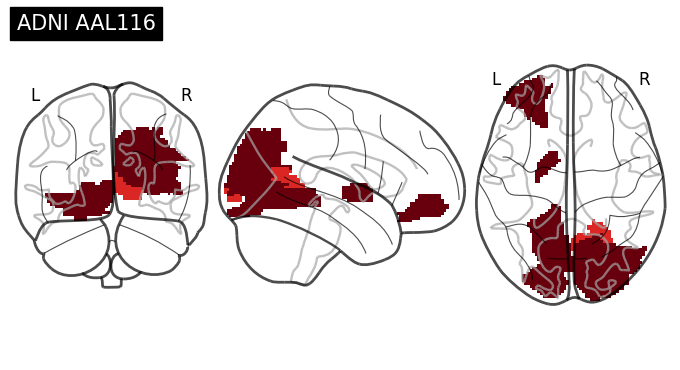}
\includegraphics[width=0.34\textwidth]{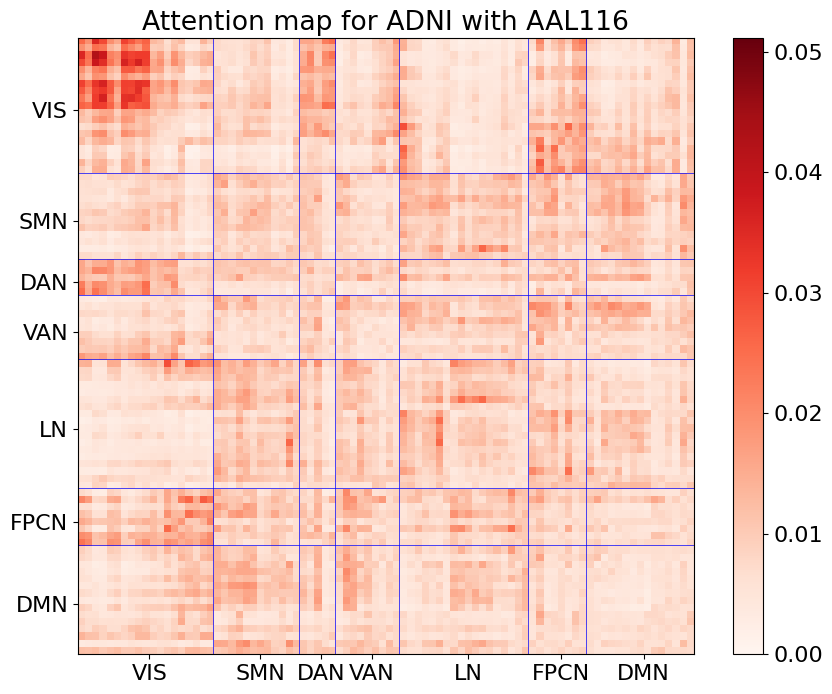}
\caption{Visualization for attention maps on ADNI. VIS = visual network; SMN = somatomotor network; DAN = dorsal attention network; VAN = ventral attention network; LN = limbic network; FPCN = frontoparietal control network; DMN = default mode network.}
\label{fig:saliency_map_adni}
\end{figure*}

Besides Schaefer100 and AAL116, we also conduct experiments with 3 atlases on ABIDE by using BASC122 \cite{BELLEC20101126}). When extending AIDFusion to scenarios involving three atlases, each component of our framework generalizes naturally from the two-atlas case. Specifically, inter-atlas message-passing is applied independently to each pair of atlases. For three atlases (e.g., A, B, and C), we construct inter-atlas edges and perform message-passing for the pairs (A, B), (A, C), and (B, C). Similarly, the subject-level and population-level consistency losses are computed for each atlas pair and then averaged across all combinations to form the final loss terms. Results shown in Table \ref{tab:atlas_results_basc} demonstrate that for five multi-atlas methods, three of them (MGRL, METAFormer, and AIDFusion) achieve better performance with three atlases compared to any two-atlas combinations. Importantly, our proposed AIDFusion still outperforms all baselines in these settings. This reinforces our claim that AIDFusion effectively integrates multi-atlas information while remaining robust to the inclusion of additional atlases. Besides, Table \ref{tab:atlas_results_basc} also shows METAFormer exhibiting a large performance gain when incorporating three atlases. However, this gain is primarily due to its relatively low performance in the two-atlas setting, rather than an indication of superior scalability. In fact, even with three atlases, METAFormer still underperforms compared to AIDFusion with only two atlases. This further highlights the effectiveness of our model in extracting comprehensive subject-level representations without requiring excessive atlas redundancy.

\subsection{Ablation Study}

To inspect the effect of the key components in AIDFusion, we conduct experiments by disabling each of them without modifying other settings. The results on ADNI dataset are reported in Table \ref{tab:ablation}. For inter-atlas message-passing (denoted as ``IA-MP'' in the table), subject-level consistency (SC) and population-level consistency (PC), we disable them by simply removing these modules. When disabling ``Disen TF'', we replace the disentangle Transformer and the identity embedding with a vanilla Transformer backbone (denoted as ``TF'' in the table). \jiaxing{We further conduct experiments by disabling the identity embedding (shown as ``Disen TF w/o ID''), and the orthogonal loss (shown as ``Disen TF w/o $L_{orth}$'') described in Section \ref{subsec:disen_tf}.} When disabling all key components (the first row in the table), our model will degenerate to MGT in Table \ref{tab:main_results}. The results demonstrate that AIDFusion with all important modules enabled achieves the best performance. The component that affects the performance most is the population-level consistency. Besides, all variants of the proposed AIDFusion outperform the MGT baseline, demonstrating the effectiveness of our model design. \revise{This table highlights not only the additive value of each module but also the superiority of our framework compared to existing multi-atlas models, even when some components are removed.}

\begin{table}[h]
\centering
\caption{Ablation study on the key components of AIDFusion on ADNI over 10-fold-CV of a single run, with the best result \textbf{bold}.}
\begin{tabular}{cccc|c}
\hline
Backbone & IA-MP & SC & PC & acc ± std    \\ \hline
TF       &                &             &  
            & 63.99 ± 4.34 \\
TF       & \checkmark     & \checkmark  & \checkmark  & 66.82 ± 1.25 \\
\jiaxing{Disen TF w/o ID}       & \jiaxing{\checkmark}     & \jiaxing{\checkmark}  & \jiaxing{\checkmark}  & \jiaxing{66.97 ± 1.35} \\
\jiaxing{Disen TF w/o $\mathcal{L}_{orth}$}       & \jiaxing{\checkmark}     & \jiaxing{\checkmark}  & \jiaxing{\checkmark}  & \jiaxing{66.06 ± 1.17} \\
Disen TF &       & \checkmark  & \checkmark  & 66.58 ± 1.72 \\
Disen TF & \checkmark     &    & \checkmark  & 66.37 ± 1.56 \\
Disen TF & \checkmark     & \checkmark  &    & 65.91 ± 2.08 \\
Disen TF & \checkmark     & \checkmark  & \checkmark  & \textbf{67.57} ± 2.04 \\ \hline
\end{tabular}
\label{tab:ablation}
\end{table}

\begin{figure*}[h]
\centering
\includegraphics[width=0.37\textwidth]{img/attention_map_adni_schaefer100.png}
\includegraphics[width=0.39\textwidth]{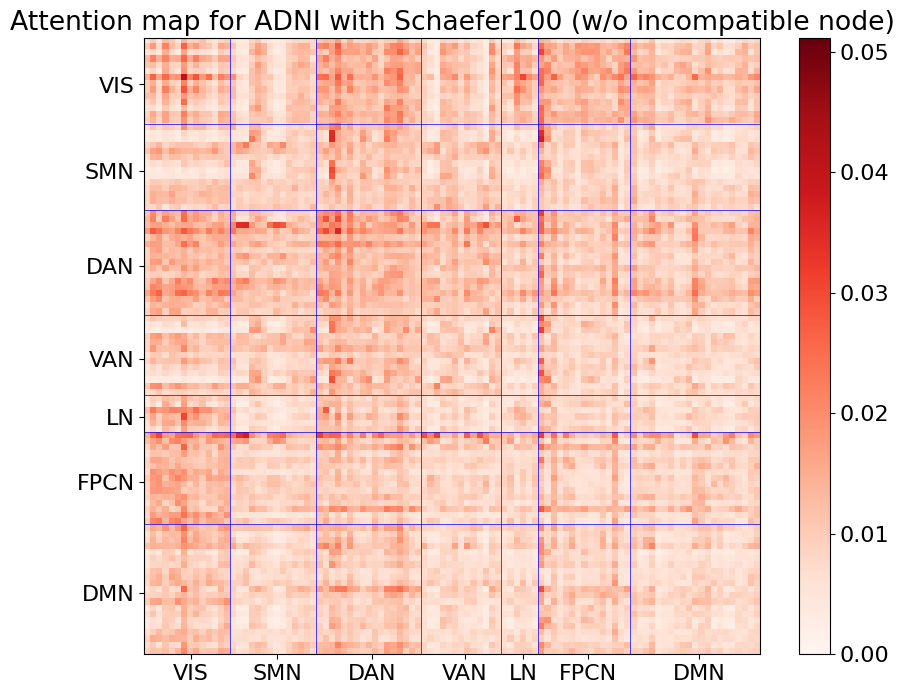}
\includegraphics[width=0.37\textwidth]{img/attention_map_adni_aal116.png}
\includegraphics[width=0.39\textwidth]{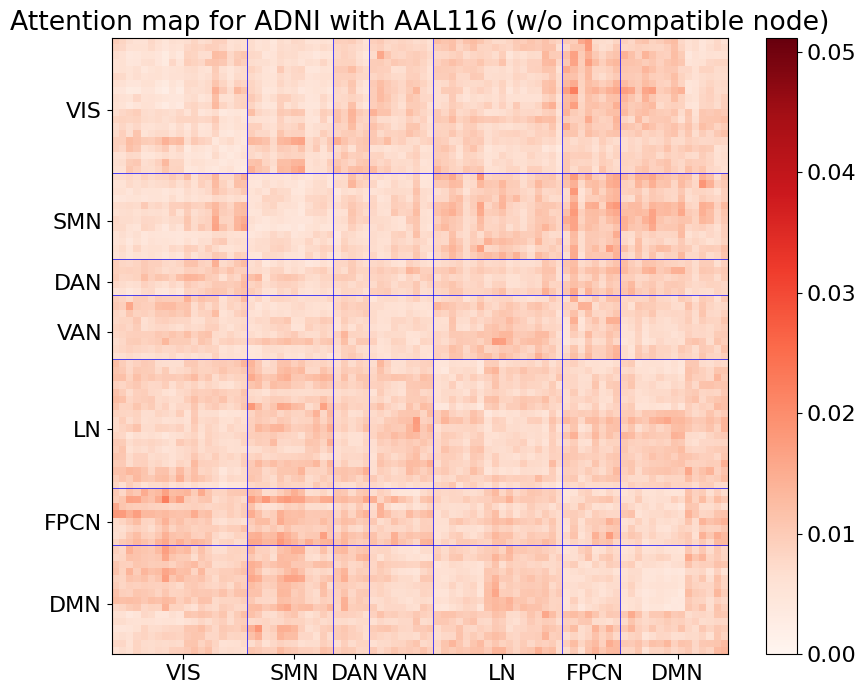}
\caption{Visualization for attention maps of AIDFusion w/ and w/o incompatible nodes.}
\label{fig:saliency_map_adni_ablation}
\end{figure*}

\subsection{Model Interpretation}

\noindent
\textbf{Biomarker Detection.} In neurodegenerative disorder diagnosis, identifying salient ROIs/connections associated with predictions as potential biomarkers is crucial. In this study, we utilize attention scores from the Transformer layer to generate heat maps for brain networks to interpret our model. We visualize these attention maps using the Nilearn toolbox~\cite{abraham2014machine}. Figure \ref{fig:saliency_map_adni} presents attention maps for two atlases, where higher attention values mean better classification potential for AD (from the ADNI dataset). We utilized 7 networks~\cite{yeo2011organization} to assess the connections between our highlighted ROIs and major networks potentially involved with disorders. ROIs from the AAL that do not overlap with these seven networks are excluded from the heat maps. The top 10 ROIs with the highest attention values are displayed in the brain view. As depicted in the attention maps, attention maps of both Schaefer and AAL atlases identify common connections between the visual network (VIS) and the dorsal attention network (DAN), recognized as key connectivities in AD research~\cite{brier2012loss,agosta2012resting}. Additionally, atlas-specific connections are highlighted. 
For example, the attention map of Schaefer atlas emphasizes connections within the default mode network (DMN) corresponding with the observations of Damoiseaux et al. \cite{damoiseaux2012functional}.
Findings on the attention map on AAL are consistent with Agosta et al. \cite{agosta2012resting}, showing that AD is associated with connectivities in VIS, especially in frontal networks. These findings suggest that AIDFusion effectively captures complementary information from different atlases. Besides, we also find some of our findings diverge from conventional neuroscientific understanding. For example, connections between VIS and somatomotor network (SMN) have a high attention weight in AIDFusion on Schaefer atlas, which may imply AD is related to the function of defining the targets of actions and providing feedback for visual activation. This insight has not been identified by existing literature in the domain of neuroscience. 
We also note that the attention map over AAL116 reveals that the cerebellar regions (the lower portion of the VIS network) receive noticeably lower attention weights compared to cortical ROIs. This observation suggests that AIDFusion is capable of downweighting atlas-specific regions that lack consistent correspondence across atlases (e.g., cerebellum absent in Schaefer100), thereby preventing spurious feature alignment. It highlights the model’s ability to selectively focus on semantically meaningful and cross-atlas-consistent regions during representation learning.

\noindent
\textbf{Case Study for Incompatible Nodes.} To further support the function of incompatible nodes, we conducted a case study comparing attention maps of AIDFusion with and without the incompatible nodes. The attention maps are shown in Figure \ref{fig:saliency_map_adni_ablation}. We can observe that, when not using incompatible nodes, the attentions of two atlases (in the right column) are remarkably imbalanced. Attention on Schaefer is much higher than that on AAL. Besides, the attention map of AAL exhibits over-smoothing and no highlighted network is found, which indicates the model is not able to extract the distinguishable connections. This case study demonstrates that the incompatible nodes enable the model to filter out the inconsistent atlas-specific information.

\begin{table*}[h]
\centering
\caption{Brain Network Classification Results (Average Accuracy ± Standard Deviation) over 10-fold-CV under out-of-distribution setting. The best results are highlighted in \textbf{bold}.}
\begin{tabular}{c|ccccc}
\hline
model      & Accuracy          & Precision          & Recall           & micro-F1            & ROC-AUC          \\ \hline
MGRL       & 62.53 ± 4.20 & 60.37 ± 4.52 & 62.27 ± 7.60  & 61.13 ± 5.25  & 62.49 ± 4.25 \\
MGT        & 56.78 ± 3.99 & 55.69 ± 5.08 & 49.40 ± 8.63 & 51.73 ± 6.17 & 56.43 ± 3.93 \\
METAFormer & 61.17 ± 2.92 & 58.71 ± 3.21 & 62.70 ± 8.13  & 60.37 ± 4.48  & 61.23 ± 3.00 \\
LeeNet     & 59.51 ± 3.61 & \textbf{62.06} ± 5.20 & 39.36 ± 15.32 & 46.21 ± 12.83 & 58.58 ± 4.05 \\
AIDFusion (ours) & \textbf{63.50} ± 3.72 & 59.67 ± 3.22 & \textbf{72.55} ± 6.13  & \textbf{65.37} ± 3.69  & \textbf{63.91} ± 3.74 \\ \hline
\end{tabular}
\label{tab:ood}
\end{table*}

\begin{table*}[h]
\centering
\caption{Time efficiency analysis. Total time (h) was recorded with a single run (including training, validation, and test) with 10-fold-CV.}
\begin{tabular}{c|cc|cc|cc|cc|c}
\hline
 & \multicolumn{2}{c|}{ABIDE} & \multicolumn{2}{c|}{ADNI} & \multicolumn{2}{c|}{PPMI} & \multicolumn{2}{c|}{Mātai} & \multirow{2}{*}{\#Param} \\
           & Time (h) & \#Epoch     & Time (h) & \#Epoch     & Time (h) & \#Epoch      & Time (h) & \#Epoch      &       \\ \hline
MGRL       & 0.56     & 261.9 \scriptsize{± 0.7} & 0.91     & 262.7 \scriptsize{± 0.8} & 0.15     & 272.2 \scriptsize{± 8.0}  & 0.09     & 291.2 \scriptsize{± 14.7} & 378k  \\
MGT        & 0.78     & 263.9 \scriptsize{± 2.4} & 0.89     & 108.2 \scriptsize{± 1.1} & 0.08     & 134.3 \scriptsize{± 10.9} & 0.18     & 266.4 \scriptsize{± 4.2}  & 273k  \\
METAFormer & 1.73     & 263.2 \scriptsize{± 1.5} & 1.47     & 268.3 \scriptsize{± 2.2} & 0.16     & 266.5 \scriptsize{± 4.7}  & 0.12     & 270.4 \scriptsize{± 3.6}  & 1886k \\
LeeNet     & 1.56     & 200.0 \scriptsize{± 0.0} & 1.76     & 200.0 \scriptsize{± 0.0} & 0.19     & 200.0 \scriptsize{± 0.0}  & \textbf{0.07}     & 200.0 \scriptsize{± 0.0}  & 526k  \\
AIDFusion (ours) & \textbf{0.12}     & \textbf{48.5} \scriptsize{± 19.0} & \textbf{0.26}     & \textbf{64.6} \scriptsize{± 14.1} & \textbf{0.03}     & \textbf{34.9} \scriptsize{± 12.0}  & 0.10     & \textbf{119.5} \scriptsize{± 13.8} & \textbf{235k}  \\ \hline
\end{tabular}
\label{tab:time}
\end{table*}

\subsection{Generalization Ability}
\label{subsec:generalize}

While task-specific biomarkers are valuable for identifying disease-relevant features, it is crucial to determine whether these biomarkers are invariant over the entire population, i.e., whether they generalize well across sub-populations and other diverse populations. To assess the generalization ability of AIDFusion, we conduct evaluations on subjects from previously unseen sites. Specifically, we have conducted additional experiments under an out-of-domain (OOD) setting following the protocol proposed in \cite{xu2025brainood}, where one site is held out as the OOD test set for each fold. This simulates the realistic scenario of deploying the model to an unseen clinical center.

We report the performance of AIDFusion and all multi-atlas baselines on ABIDE dataset, in which the test set includes both in-domain (ID) and OOD subjects. As shown in Table~\ref{tab:ood}, AIDFusion consistently achieves superior performance compared to other multi-atlas methods. This demonstrates the robustness and transferability of our model across sites with heterogeneous data distributions.

\subsection{Time Efficiency}

We conducted an experiment to compare the total runtime cost of AIDFusion with other multi-atlas baselines. \jiaxing{Note that our method requires computing population-level consistency, which scales with the number of subjects rather than being a linear per-subject cost.} The results, reported in Table \ref{tab:time}, demonstrate that AIDFusion requires dramatically fewer epochs to converge, resulting in significantly less time spent on ABIDE, ADNI, and PPMI datasets. For the Mātai dataset, AIDFusion's time cost is still comparable with the other baselines. Besides, since AIDFusion does not contain any repeat layers as other baselines do, it has fewer parameters and thus results in higher efficiency. This showcases the efficiency of the proposed AIDFusion.

\subsection{Hyperparameter Analysis}
\label{subsec:hyperparams}

We study the sensitivity of four trade-off hyperparameters in Eq. (\ref{eq:total_loss}). All experiments are conducted on the ADNI dataset. We tune the value of $\lambda_1$ from 1e0 to 1e2, $\lambda_2$ from 1e0 to 1e2, $\lambda_3$ from 1e-6 to 1e-4 and $\lambda_4$ from 1e-1 to 1e1. The results presented in Table \ref{tab:tune_lambda} show that our model performs the best when $\lambda_1=$ 1e1, $\lambda_2=$ 1e1, $\lambda_3=$ 1e-5 and $\lambda_4=$ 1e0. We can exhibit that these trade-off hyperparameters in the loss function will marginally affect the model performance on ADNI (less than 1\%), which demonstrates the stability of AIDFusion.

\begin{table}[h]
\caption{The hyperparameter sensitivity analysis for AIDFusion on ADNI dataset over 10-fold-CV of a single run.}
\centering
\begin{tabular}{cccc|c}
\hline
$\lambda_1$  & $\lambda_2$  & $\lambda_3$   & $\lambda_4$   & acc ± std    \\ \hline
1e0 & 1e1 & 1e-5 & 1e0  & 66.97 ± 1.95 \\
1e1 & 1e0 & 1e-5 & 1e0  & 66.44 ± 2.88 \\
1e1 & 1e1 & 1e-6 & 1e0  & 67.04 ± 2.20 \\
1e1 & 1e1 & 1e-5 & 1e-1 & 66.82 ± 1.98 \\
1e1 & 1e1 & 1e-5 & 1e0  & \textbf{67.57} ± 2.04 \\
1e1 & 1e1 & 1e-5 & 1e1  & 66.82 ± 2.64 \\
1e1 & 1e1 & 1e-4 & 1e0  & 67.04 ± 2.21 \\
1e1 & 1e2 & 1e-5 & 1e0  & 66.89 ± 2.10 \\
1e2 & 1e1 & 1e-5 & 1e0  & 66.21 ± 2.34 \\ \hline
\end{tabular}
\label{tab:tune_lambda}
\end{table}

\section{Conclusion}
\label{sec:conclusion}

In this paper, we presented the Atlas-Integrated Distillation and Fusion network (AIDFusion), a novel approach to multi-atlas brain network classification. The disentangle Transformer mechanism, combined with inter-atlas message-passing and consistency constraints, effectively integrates complementary information across different atlases and ensures cross-atlas consistency at both the subject and population levels. Our extensive experiments on four fMRI datasets demonstrate that AIDFusion outperforms state-of-the-art methods in terms of classification accuracy and efficiency. Moreover, the patterns identified by AIDFusion align well with existing domain knowledge, showcasing the model's potential for providing interpretable insights into neurological disorders. 
We hope our work inspires further research in multi-atlas brain network analysis and demonstrates its significance in real-world applications, such as early diagnosis and personalized treatments for neurodegenerative diseases.

\section*{Acknowledgments}
This research/project is supported by the Ministry of Education, Singapore under its Academic Research Fund Tier 1 (RG16/24). Any opinions, findings and conclusions or recommendations expressed in this material are those of the author(s) and do not reflect the views of the Ministry of Education, Singapore.



 
%
\bibliographystyle{IEEEtran}
\bibliography{citation}

\end{document}